\pdfoutput=1
\documentclass[notitlepage,twocolumn,prl]{revtex4-1}

\usepackage{amsmath}
\usepackage{changes}
\usepackage{graphicx}

\begin{document}

\title{Inter-mode breather solitons in optical microresonators}

\author{Hairun Guo}
\author{Erwan Lucas}
\author{Martin H. P. Pfeiffer}
\author{Maxim Karpov}
\author{Miles Anderson}
\author{Junqiu Liu}
\author{Michael Geiselmann}
\author{John D. Jost}
\author{Tobias J. Kippenberg}
\email{tobias.kippenberg@epfl.ch}
\affiliation{{\'E}cole Polytechnique F{\'e}d{\'e}rale de Lausanne (EPFL), Lausanne, CH-1015, Switzerland}

\date{\today}

\begin{abstract}
  The observation of temporal dissipative Kerr solitons in optical microresonators provides, on the applied side, compact sources of coherent optical frequency combs that have already been applied in coherent communications, dual comb spectroscopy and metrology. On a fundamental level, it enables the study of soliton physics in driven nonlinear cavities. 
  Microresonators are commonly multimode and, as a result, inter-mode interactions inherently occur among mode families -- a condition commonly referred to as ``avoided mode crossings''. Avoided mode crossings can cause soliton decay, but can also modify the soliton spectrum, leading to e.g. the formation of dispersive wave like power-enhanced spectral elements, and inducing a spectral recoil. Yet, to date, the entailing temporal soliton dynamics from inter-mode interactions has rarely been studied, but is critical to understand regimes of soliton-stability. Here we report the discovery of an inter-mode breather soliton -- soliton featuring breathing from inter-mode interactions. Such breathing dynamics occurs within a laser detuning range where conventionally stationary (i.e. stable) dissipative solitons are expected. We demonstrate experimentally the phenomenon in two microresonator platforms (crystalline magnesium fluoride and photonic chip-based silicon nitride microresonators), and theoretically describe the dynamics based on a pair of coupled Lugiato-Lefever equations. We demonstrate experimentally that the breathing is associated with a periodic energy exchange between the soliton and another optical mode family. We further show that inter-mode interactions can be modeled by a response function acting on dissipative solitons. The observation of breathing dynamics in the conventionally stable soliton regime is critical to applications, ranging from low-noise microwave generation, frequency synthesis to spectroscopy. On a fundamental level, our results provide new understandings of the rich dissipative soliton dynamics in multimode nonlinear cavities.
\end{abstract}

\maketitle

\section{Introduction}
The generation of dissipative Kerr solitons (DKS) in high-\emph{Q} optical microresonators \cite{herr2014temporal,brasch2016photonic} enables fully coherent optical frequency combs with microwave repetition rate, large bandwidth and chip-scale compactness, and is physically based on the double balance between dispersion and nonlinearity, as well as dissipation and a driving source.
Such soliton-based frequency combs have been demonstrated in a wide variety of microresonator platforms \cite{yi2015soliton,liang2015high,wang2016intracavity,joshi2016thermally} and already been used in applications, ranging from terabit coherent communications \cite{marin2016microresonator}, dual comb spectroscopy \cite{suh2016microresonator,dutt2016chip} and phase coherent microwave-to-optical links \cite{jost2015counting,brasch2017self}.
Furthermore, they also represent a new platform for studying the rich dissipative soliton physics \cite{akhmediev2008dissipative}.
Beyond the study of stable soliton states, oscillatory or periodic soliton evolutions in a nonlinear cavity have attracted increasing attention. Dynamics, ranging from pulsating solitons in mode-locked fiber lasers \cite{bao2015observation}, soliton molecular vibrations in Kerr-lens mode-locked lasers \cite{herink2017real}, and temporal breathing solitons in dissipative Kerr cavities \cite{leo2013dynamics}, have been experimentally observed and characterized, which reveal complex yet self-consistent soliton regimes. The periodic evolution also implies the property of topological protection in such cavity solitons \cite{herink2017real}. In particular, breathing DKS -- soliton featuring periodic oscillations in both amplitude and duration -- were recently experimentally investigated in several Kerr microresonator platforms \cite{bao2016observation,yu2016breather,lucas2016breathing}. In these studies, the breathing phenomenon corresponds to an intrinsic dynamical instability of dissipative Kerr cavity systems \cite{barashenkov1996existence,leo2013dynamics,parra2014dynamics,godey2014stability} described by a standard Lugiato-Lefever equation (LLE) model (or equivalently a set of coupled mode equations) \cite{lugiato1987spatial,chembo2010modal,coen2013modeling,chembo2013spatio}, which exists near the low-detuning boundary of the soliton existence domain. For stable single and multiple soliton states, the detuning dependent existence domain has also been experimentally analyzed and compared to the theory \cite{lucas2016study}, and is modified by the thermal effect \cite{guo2016universal}.


To date, most microresonator platforms are \emph{multimode} either inherently or on grounds to achieve the proper waveguide dispersion. For instance, to support DKS, thick silicon nitride (${\rm Si_3N_4}$) waveguide-based microresonators \cite{riemensberger2012dispersion,pfeiffer2016photonic} are required to obtain the necessary anomalous group velocity dispersion (GVD) and are therefore prone to having inter-mode coupling. When two modes exhibit a similar resonance frequency and are coupled, a pair of symmetric and asymmetric modes is formed \cite{carmon2008static}, a situation commonly referred to avoided mode crossing (abbreviated as AMX hereafter).
Previous studies revealed that strong and multiple AMX-induced deviations in the dispersion can suppress the soliton formation \cite{herr2014mode}, and can be mitigated by introducing a mode filtering section in integrated microresonators \cite{kordts2016higher}. In cases with only few or weak AMXs, DKS can still form but will typically show a localized power change in the spectrum of soliton-based frequency comb \cite{herr2014mode}, implying a deviation of the soliton waveform from an analytical ansatz of the standard LLE \cite{barashenkov1996existence,wabnitz1993suppression,herr2014temporal}. In particular, this can equivalently lead to the formation of dispersive waves and introduce a soliton spectral recoil \cite{matsko2016optical,yang2016spatial,yi2016single}. In addition, AMX has also been shown to allow Kerr frequency combs in the normal dispersion regime \cite{xue2015normal,miller2015tunable,soltani2016enabling}.
Moreover, a recent study \cite{yi2016single} showed that an AMX-induced single-mode dispersive wave can reduce the repetition rate noise and lead to a noise ``quiet'' point, by a mechanism in which the detuning dependent soliton repetition rate due to the Raman self-frequency shift \cite{karpov2016raman,yi2016theory} is compensated by the spectral recoil induced by the dispersive wave. Yet to date, little is known about the temporal soliton dynamics in the presence of inter-mode interactions.

\begin{figure*}[t!]
  \centering{
  \includegraphics[width = 1 \linewidth]{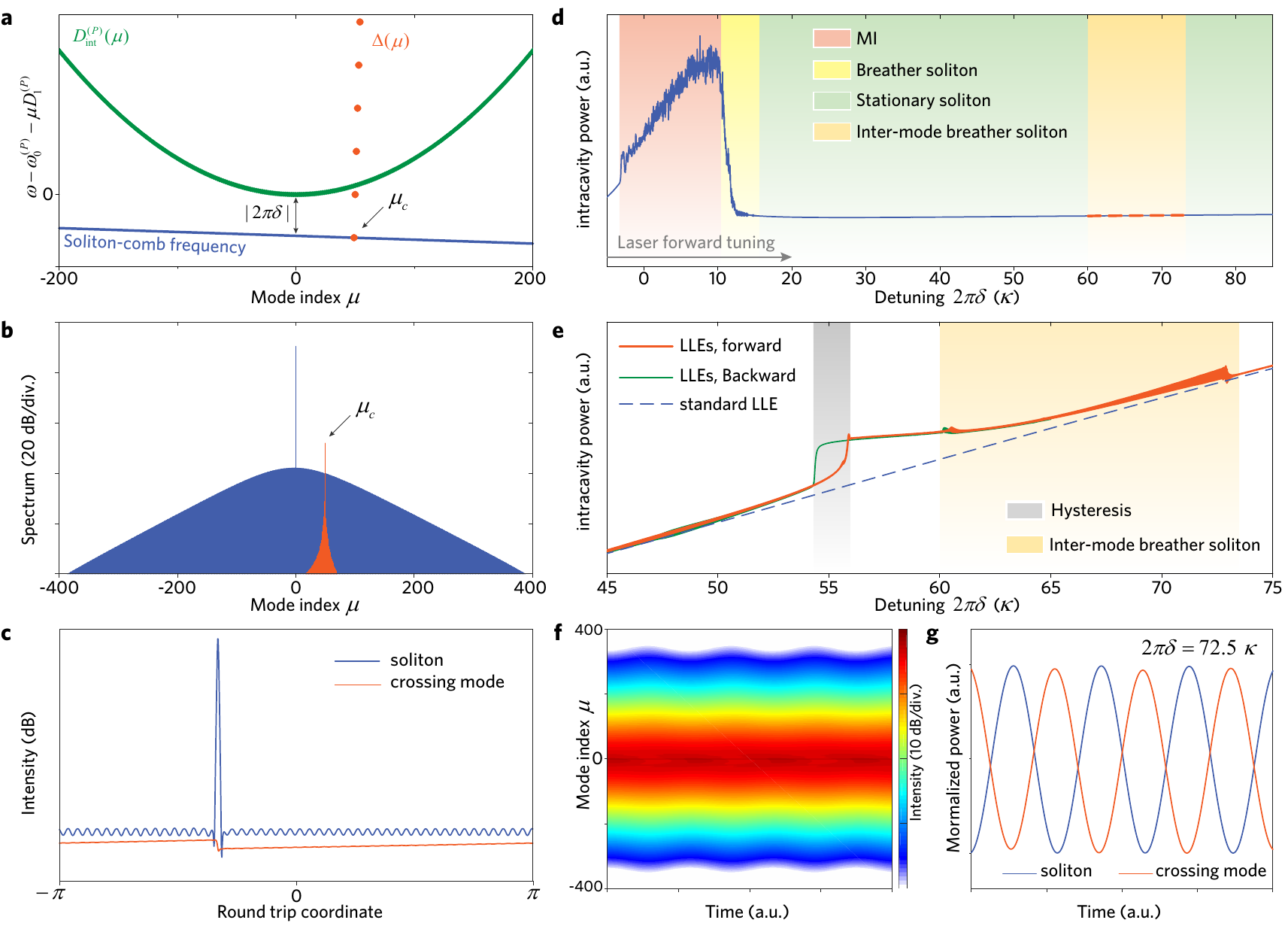}
  }
  \caption{\textbf{Numerical simulation of inter-mode breather soliton in a Kerr microresonator.}
  (a) Integrated dispersion of a (primary) soliton-supporting mode family (green dots) in which the parabolic profile indicates the anomalous dispersion in this mode family. In the same frame, the crossing mode family shows a slop profile corresponding to ${\Delta(\mu)}$ (red dots). Once the soliton-based frequency comb is formed with a detuning (${\delta}$), the soliton-comb frequency represents nearly a constant (${ -2\pi\delta}$) (blue line), implying the equal distance between comb teeth, where the slight slope corresponding to a change in the FSR (compared to ${D_1^{(P)}/{2\pi}}$) is caused by the soliton central frequency shift that is either by dispersive wave induced soliton spectral recoil or by the Raman self-frequency shift. Thus, the phase matching between the soliton and the wave in the crossing mode (${\mu_c}$) is ${\Delta(\mu_c)+2\pi\delta \approx 0}$. (b) Simulated single-soliton-based frequency comb in the primary mode family (blue lines) and a narrow-band waveform in the crossing mode family (orange lines). (c) Intracavity field patterns in both mode families. (d) Intracavity power trace over the laser detuning, based on a standard LLE model in the absence of inter-mode interactions (blue line). Three stages, modulation instability (MI) regime (red area), breather soliton (yellow area) and stationary soliton (green area), are marked as typical transitions of frequency comb states in the generation process. The inter-mode breather soliton exists in the region where usually stationary DKS are expected (orange area). (e) DKS in the presence of inter-mode interactions (based on the coupled LLEs model) show a different behavior, including a hysteresic power jump (gray area) and an oscillatory behaviour (orange area). (f) Spectral envelope evolution of a single inter-mode breather soliton. (g) Out-of-phase oscillations in the energy of the soliton and the crossing mode waveform.
  \label{fig_simu}
}
\end{figure*}


In this letter, we report a novel observation, showing that AMX can lead to \emph{breather solitons}. We demonstrate that the breathing dynamics is due to a periodic energy exchange between the cavity soliton and a crossing mode family, which is fundamentally distinct from previously-observed breathers \cite{bao2016observation,yu2016breather,lucas2016breathing}. Therefore we refer to this effect as an \emph{inter-mode breather soliton}. Importantly, the inter-mode breather soliton occurs within a laser detuning range where \emph{stationary} (i.e. stable) DKS are usually expected. We observe this unexpected dynamics in two of our microresonator platforms, i.e. a magnesium fluoride (${\rm MgF_2}$) crystalline resonator and a ${\rm Si_3N_4}$ waveguide-based microresonator. Such an \emph{inter-mode breathing} dynamics is also confirmed by numerical simulations, based on a set of coupled LLEs \cite{yi2016single,daguanno2016nonlinear}.
Physically, inter-mode interactions are understood as a response effect acting on cavity DKS, such that a single LLE-like model is derived to describe the soliton dynamics, which can fully account for the inter-mode breathing. From an application perspective, understanding and avoiding inter-mode breathing is critical for applications ranging from self-referenced soliton combs \cite{jost2015counting} for frequency metrology to the generation of low-noise microwaves \cite{liang2015high}.

\section{Theory and simulation}
Inter-mode interactions among microresonator mode families consist of both linear and nonlinear couplings. The linear coupling can be described by coupled mode theory \cite{haus1991coupled,wiersig2006formation}, while the nonlinear coupling mainly refers to the cross-phase modulation in Kerr cavities. In a weak-coupling limit, one can assume that DKS remain supported in a primary (``${\mathcal P}$'') mode family that is linearly coupled to a second, crossing mode family (``${\mathcal C}$'').
The integrated dispersion of the primary soliton-supporting mode family is defined as ${D_{\rm int}^{({\mathcal P})}(\mu) = \omega_{\mu}^{({\mathcal P})} - \omega_{0}^{({\mathcal P})} - \mu \cdot D_{1}^{({\mathcal P})}}$ (where ${\omega_{\mu}^{({\mathcal P})}}$ indicates the resonance frequency over the relative mode index ${\mu}$, ${\mu = 0}$ is the central pumped mode, and ${D_{1}^{({\mathcal P})}/{2\pi}}$ is the free spectral range, FSR), whereas relative to this frequency grid,
the crossing mode family has the frequency given by ${\Delta(\mu) = \omega _{0}^{({\mathcal C})} - \omega _{0}^{({\mathcal P})} + \mu (D_{1}^{({\mathcal C})} - D_{1}^{({\mathcal P})})}$ (Fig. \ref{fig_simu}(a)). Therefore, an AMX occurs around the mode where ${D_{\rm int}^{({\mathcal P})}(\mu) \approx \Delta(\mu)}$.
Moreover, a soliton-based frequency comb is fully coherent with equally spaced spectral elements (i.e. \emph{dispersionless}), such that it appears as a straight line in the frame of ${D_{\rm int}^{({\mathcal P})}(\mu)}$, see Fig. \ref{fig_simu}(a), and is supported in a continuous range of laser detuning ${2\pi\delta = \omega _{0}^{({\mathcal P})} - \omega_p > 0}$ (i.e. the pump (${\omega _p}$) is necessarily red-detuned to the central mode).
Therefore, at a given detuning ${{\delta}}$, the phase matching condition between the cavity soliton to a wave in the crossing mode ${\mu _c}$ is given by:
\begin{equation}
2\pi\delta + \Delta(\mu_c) \approx 0
\label{eq_pm}
\end{equation}
%

The dynamics of DKS in the presence of inter-mode interactions can be fully explained by two sets of coupled mode equations (or equivalently two coupled LLEs) including both the linear coupling and the cross-phase modulation \cite{yi2016single,daguanno2016nonlinear}, as detailed in the supplementary information (SI).
Alternatively, effects of the linear coupling can be approximated by a suitable response incorporated into a single set of coupled mode equations (or a single LLE framework) that accounts for soliton dynamics in the primary mode family. In a ``co-traveling'' frame (i.e. a frame that is co-traveling with the soliton waveform centered at the pump frequency) such a single LLE-like model is written as:
\begin{multline}
\label{eq_lle_like}
\frac{\partial {\tilde A}_{\mu}^{({\mathcal P})}(t)}{\partial t} = \left( { - \frac{\kappa^{({\mathcal P})}}{2} + i(2\pi\delta) + i D_{\rm int}^{({\mathcal P})}(\mu) } \right){\tilde A}_{\mu}^{({\mathcal P})} \\
+{\tilde R}_{\rm c}(\mu) {\tilde A}_{\mu}^{(P)} - i g {\mathcal F} \left[ {|A^{(P)}|^2 A^{(P)}} \right]_{\mu} + \delta '_{\mu 0}{\sqrt{\kappa _{\rm ex}} \cdot s_{\rm in}}
\end{multline}
where ${{\tilde A}_{\mu}^{({\mathcal P})}}$ and ${{A^{({\mathcal P})}}}$ are the spectral and temporal envelopes of DKS, respectively (related via ${A^{({\mathcal P})}(t)=\sum_{\mu}{\tilde A}_{\mu}^{({\mathcal P})} e^{-i \mu D_1^{({\mathcal P})} t}}$), ${\kappa^{({\mathcal P})}}$ is the loss rate of the soliton-supporting primary mode family, ${g}$ is the single photon induced Kerr frequency shift (see SI for the full definition), ${\kappa_{\rm ex}}$ is the coupling rate and ${|s_{\rm in}|^2}$ denotes the pump power, ${\delta '_{\mu 0}}$ is the Kronecker delta, and ${{\mathcal F}[~]_{\mu}}$ represents the $\mu$'th frequency component of the Fourier series.
We derive an \emph{inter-mode response} accounting for the linear coupling:
\begin{equation}
{\tilde R}_{\rm c}(\mu) = \frac{G^2}{{\mathcal D}(\mu)} \left[e^{({\mathcal D}(\mu) \cdot t)} + 1 \right]
\label{eq_rc}
\end{equation}
where ${{\mathcal D}(\mu) = -\frac{\kappa^{({\mathcal C})}}{2} + i(2\pi\delta + \Delta(\mu))}$ contains the phase matching condition.
The response bandwidth is defined by the loss rate of the crossing mode family (${\kappa^{({\mathcal C})}}$) and the response amplitude is scaled by the strength of the linear coupling (${G}$) between the primary and the crossing mode families.
Compared to the full model, such a single LLE-like model can equivalently reproduce the soliton dynamics in the presence of linear inter-mode interactions.
Nevertheless, if there exists more than one crossing mode fulfilling the phase matching condition (Eq. \ref{eq_pm}), the LLE-like model can be flexibly extended by including several inter-mode responses (Eq. \ref{eq_rc}).

\begin{figure}[t!]
\centering{
\includegraphics[width = 1 \linewidth]{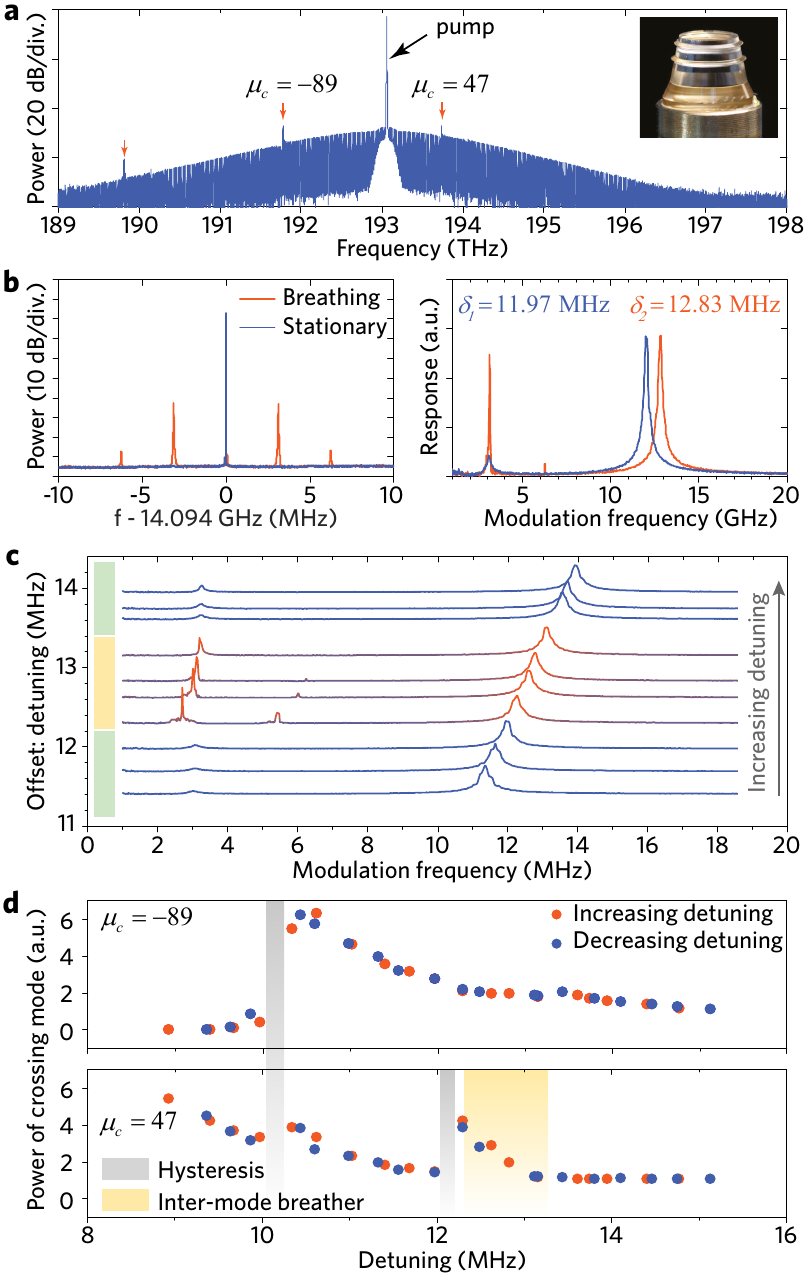}
}
\caption{\textbf{Observation of an inter-mode breather soliton in a ${\rm \bf MgF_2}$ crystalline microresonator} (${\sim 14 ~{\rm GHz}}$ FSR). (a) Experimentally generated single-soliton-based frequency comb exhibiting spikes (i.e. enhanced power in comb teeth) from inter-mode interactions. (b) The beatnote measurements (left) and system's response measurements (right) for both the stationary soliton state and the inter-mode breathing state. (c) Measured evolution of the system's response when increasing the laser detuning. The soliton features breathing in the detuning range ${\rm 12.3 - 13.2 ~MHz}$) where a strong frequency tone (and harmonics) is observed in the system's response close to the {${\mathcal S}$-resonance}. (d) Measured power of single comb mode ${\mu_c = 47}$ and ${\mu_c = -89}$ for both increasing and decreasing laser detuning. In the gray area  bistable behavior is predicted, but not resolved in the experiment, attributed to a too weak coupling rate (${G}$) between the mode families.
\label{fig_mgf2}
}
\end{figure}

\begin{figure}[t!]
\centering{
\includegraphics[width = 1 \linewidth]{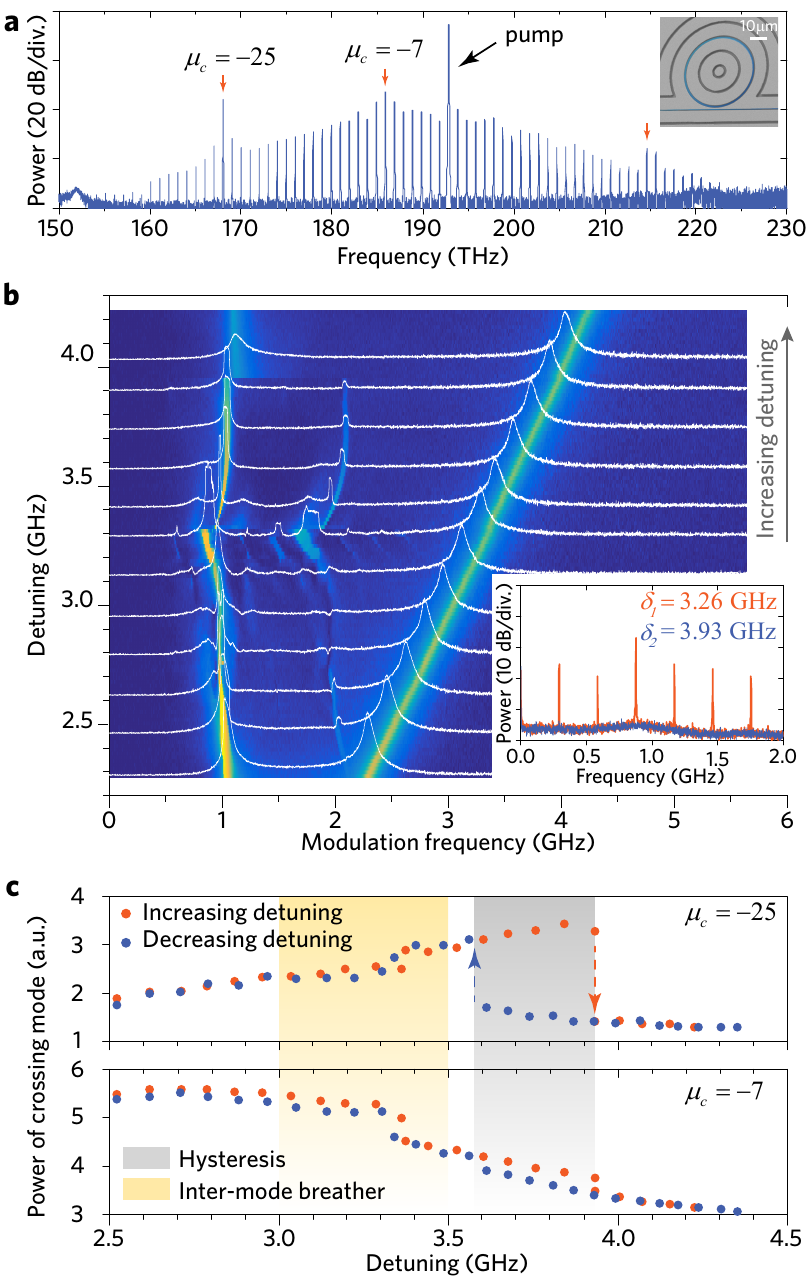}
}
\caption{\textbf{Observation of an inter-mode breather soliton in a chip-scale ${\rm \bf Si_3N_4}$ microresonator and hysteresis in the comb profile}. (a) Experimentally generated single-soliton-based frequency comb in silicon nitride resonators with (with a ${\sim 1 ~{\rm THz}}$ FSR) featuring spikes (i.e. power-enhanced comb teeths) from inter-mode interactions. (b) Measured evolution of the system's response when increasing the laser detuning. The inter-mode breathing is in the detuning range ${\rm 2.4-3.9 ~GHz}$). Inset: RF measurements for both the stationary and the breathing soliton states. (c) Measured power of single comb mode ${\mu_c = -7}$ and ${\mu_c = -25}$ for both forward and backward laser detuning. The measurements reveal a clear hysteresis of the enhanced comb teeth power (gray area).
\label{fig_si3n4}
}
\end{figure}

We next performed simulations based on the coupled LLEs.
Figure. \ref{fig_simu} shows a simulation of a single-soliton-based frequency comb in the presence of inter-mode interactions.
The soliton comb envelope in the primary mode family remains overall a ${\rm sech^2}$-profile.
In the crossing mode family a spectrally narrow-band waveform is generated, contributing a single, power-enhanced wave (in the mode ${\mu_c}$) to the overall soliton comb spectrum (Fig. \ref{fig_simu}(b)), which is phase matching to the cavity soliton.
This power-enhanced comb tooth then behaves similarly to a dispersive wave \cite{yi2016single} that causes a temporal oscillation in the intracavity field pattern (Fig. \ref{fig_simu}(c)), and induces a soliton recoil such that the comb envelope is shifted in a spectral direction opposite to that of the dispersive wave.
The temporal intracavity field in the crossing mode family is almost continuously distributed corresponding to the power-enhanced wave, but features a power step induced by the soliton via the cross-phase modulation (Fig. \ref{fig_simu}(c)).

Usually, in the absence of inter-mode interactions, DKS are known to exist within a continuous range of the laser detuning, called the soliton existence range, in which the soliton is \emph{stationary} and the soliton power smoothly evolves over the change of detuning (green shading area in Fig. \ref{fig_simu}(d)). This is because the soliton peak intensity as well as the pulse duration is scaled by the detuning (${\delta}$) \cite{Coen2013scaling, herr2014temporal}. In contrast to such a stationary soliton state, a breathing state with an oscillatory power trace also intrinsically exists in a Kerr cavity but in a narrow detuning range adjacent to the lower boundary of the soliton existence range (yellow shading area).
Here, however, we observe a strikingly different soliton dynamics \emph{within} its stationary existence range (highlighted as the orange shading area in Fig. \ref{fig_simu}(d)).
First, when tuning ${\delta}$, the intracavity power shows a sharp jump mainly contributed by the formation of the power-enhanced wave localized in the mode ${\mu_c}$ of the crossing mode family (Fig. \ref{fig_simu}(e)).
Comparing traces in both forward (increased detuning) and backward (decreased detuning) scans of the detuning, a hysteretic behavior on the soliton power is revealed, which is in agreement to a recently proposed theory \cite{yi2016single}: it is a result of the single-mode dispersive wave induced soliton spectral recoil that leads to a modification of the phase matching criterion and entails a power bistability with respect to the detuning.

Second and most interestingly, we discover a \emph{breathing} dynamics from inter-mode interactions, as indicated by oscillations and increased amplitude jitter in the power trace in close vicinity of the power jump (Fig. \ref{fig_simu}(e)), which we termed as inter-mode breather soliton.
A periodic spectrum evolution of such a breather soliton is also observed in the simulation (Fig. \ref{fig_simu}(f)), which reveals an oscillation of the soliton pulse duration -- an essential feature of breather solitons.
Moreover, a closer analysis reveals an energy exchange regime between the soliton and the waveform in the crossing mode family, where \emph{out-of-phase} power oscillations are observed (Fig. \ref{fig_simu}(g)).

\section{Experimental results}
The novel inter-mode breather solitons were experimentally obtained in two microresonator platforms, i.e. ${\rm MgF_2}$ crystalline microresonators and chip-scale ${\rm Si_3N_4}$ microresonators (cf. SI for platform details). In both platforms, a single cavity soliton can be deterministically generated in the cavity by using the laser tuning method \cite{herr2014temporal,guo2016universal}. In general, soliton-based frequency combs can be characterized, by measuring their spectral envelope, the radio frequency (RF) beatnote or the RF spectrum (for low-frequency noise characterization), and the system's response to a pump modulation \cite{guo2016universal}. In particular, the latter enables to measure the effective laser-to-resonance detuning (i.e. ${\delta}$) -- a key parameter for cavity DKS. Stabilizing the effective detuning enables us to directly compare experimentally generated DKS to the theoretical stability chart \cite{lucas2016study}, and especially in this work, identify deviations from the expected dynamics, which is induced by inter-mode interactions. The system's response also constitutes a fingerprint of the soliton state, by showing a characteristic \emph{double resonance} feature, known as the ${\mathcal C}$-resonance that indicates the effective detuning, and the soliton-related ${\mathcal S}$-resonance \cite{guo2016universal}.

Here, we focus on the soliton dynamics when the effective detuning is well within the stationary soliton existence range such that the ${\mathcal C}$- and ${\mathcal S}$-resonances are far separated. This is to avoid the intrinsic breathing dynamics which appears at a much lower detuning value \cite{lucas2016breathing}. In a ${\rm MgF_2}$ microresonator, we generated a single-soliton-based frequency comb (FSR ${\sim 14 ~{\rm GHz}}$) that has an overall ${\rm sech^2}$-shape spectral envelope, but has few power-enhanced comb teeth due to the phase matching to the cavity soliton  (Fig. \ref{fig_mgf2}(a)). While sweeping the laser frequency, such that the effective detuning is continuously changed but remains within the soliton existence range, we observed the appearance of a breathing feature (Fig. \ref{fig_mgf2}(b,c)) in the form of sidebands on the RF beatnote of the comb with a fundamental breathing frequency of ${\sim 3 ~{\rm MHz}}$. Concomitantly, in the system's response, a strong-amplitude tone appears at the same frequency and close to the ${\mathcal S}$-resonance. Such characteristics emerge in a narrow detuning range ($\delta \sim 12.8$~MHz) and are missing when the soliton is outside of this range. In experiments, we also monitored the power of each power-enhanced wave (e.g. in the modes ${\mu_c = 47}$ and ${\mu_c = -89}$) and observed an abrupt power increase of these waves over the detuning (Fig. \ref{fig_mgf2}(d)).

Similarly, in a ${\rm Si_3N_4}$ microresonator, a single-soliton-based frequency comb (with FSR ${\sim 1 ~{\rm THz}}$) was also generated (Fig. \ref{fig_si3n4}(a)) where inter-mode interactions result in a dispersive wave like spectral wave packet rather than a single, power-enhanced wave, implying that the bandwidth of the inter-mode response ${{\tilde R}_c}$ is large (compared to the comb teeth spacing). Sweeping the effective detuning in the stationary soliton existence range, frequency tones appear in both the RF spectrum and the system's response, in which a fundamental breathing frequency of ${\sim 1 ~{\rm GHz}}$ is identified (Fig. \ref{fig_si3n4}(b)). Monitoring the power in the phase matched modes as a function of laser detuning, we observed clearly the \emph{bistability}. Such a phenomenon, as previously only theoretically predicted (referred to the hysteresis in the power of the avoided mode crossing induced dispersive wave) \cite{yi2016single}, are experimentally observed here in both modes ${\mu_c = -7}$ and ${\mu_c = -25}$, when the laser tuning direction is reversed (Fig. \ref{fig_si3n4}(c)).

Interestingly, we notice that the breathing frequency of ${\sim 3 ~{\rm MHz}}$ in the ${\rm MgF_2}$ resonator and ${\sim 1 ~{\rm GHz}}$ in the ${\rm Si_3N_4}$ microresonator are both similar to that of an intrinsic breathing soliton \cite{lucas2016breathing}. In between these two type of microresonators, the three-orders of magnitude difference in the breathing frequency can be related to the difference in the resonator's \emph{Q} factors, respectively a ${{\mathcal O}(0.1)~{\rm MHz}}$ resonance linewidth in the ${\rm MgF_2}$ resonator and a ${{\mathcal O}(100)~{\rm MHz}}$ linewidth in the ${\rm Si_3N_4}$ microresonator.

\begin{figure}[t!]
\centering{
\includegraphics[width = 1 \linewidth]{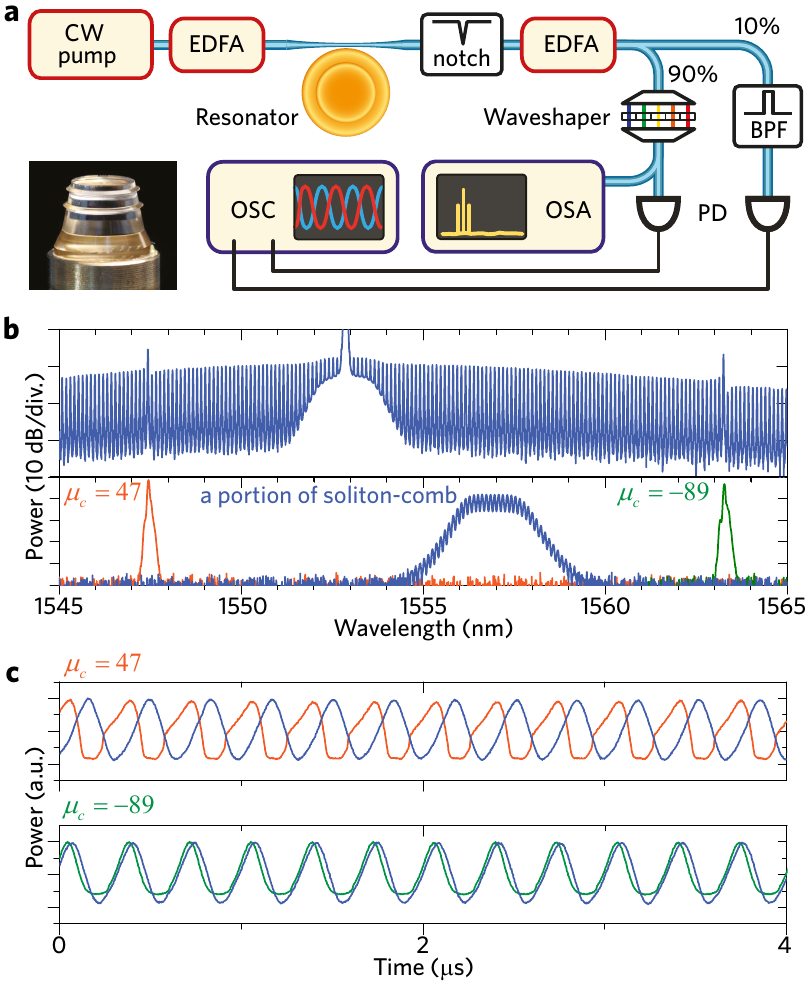}
}
\caption{\textbf{Observation of periodic energy exchange in an inter-mode breather soliton state}
(a) Schematic of the experimental setup. EDFA: erbium-doped fiber amplifier; BPF: bandpass filter; PD: photo-detector; OSC: oscilloscope; OSA: optical spectral analyzer. (b) Generated single-soliton-based frequency comb in the ${\rm MgF_2}$ microresonator (upper) and filtered components (lower) by using a waveshaper and by a bandpass filter. (c) Recorded fast power evolution of a single filtered comb mode, ${\mu_c = 47}$ (red) and ${\mu_c = -89}$ (green) compared to a ``regular'' filtered portion of the soliton spectrum (blue line). The power in ${\mu_c = 47}$ shows an out-of-phase oscillation to the soliton (upper), while the power in ${\mu_c = -89}$ shows an in-phase oscillation (lower).
\label{fig_waveshaper}
}
\end{figure}

Furthermore, we experimentally demonstrate an energy exchange regime in the inter-mode breather soliton (see Fig. \ref{fig_waveshaper}), by measuring both the power in the phase matched mode and the power of the soliton. Using a wave-shaper (operational wavelength 1527 -- 1600 nm, see sketch in Fig. \ref{fig_waveshaper}(a)), power-enhanced waves in modes ${\mu_c = 47}$ and ${\mu_c = -89}$ of the ${\rm MgF_2}$ resonator comb are selectively filtered (Fig. \ref{fig_waveshaper}(b)). Their oscillatory power traces (reflecting the breathing) are compared to that of a portion of the single-soliton-based frequency comb obtained through a bandpass filter. In this way, we observe that the power in the mode ${\mu_c = 47}$ shows an \emph{out-of-phase} oscillation with respect to the soliton power, while the power in the mode ${\mu_c = -89}$ oscillates \emph{in-phase} (see Fig. \ref{fig_waveshaper}(c)).
These observations are in excellent agreement with our simulations (Fig. \ref{fig_simu}(g)), implying that the wave mainly localized in the mode ${\mu_c = 47}$ of the crossing mode family is actively interacting with the soliton in the primary mode family, and is causing the breathing through energy exchange.
In contrast, power-enhanced waves that are in-phase oscillating do not lead to the breathing dynamics.



\section{Conclusion}
As a conclusion, we investigated a novel \emph{breathing} dynamics of cavity dissipative solitons in the presence of inter-mode interactions that originate from AMXs in multimode Kerr microresonators. AMXs lead to the formation of power-enhanced comb teeth in the soliton-based frequency comb spectrum, through a phase matching condition. The power-enhanced wave mainly localized in the crossing mode family not only exhibits a hysteresis in the power evolution over the laser detuning, but also can feature an energy exchange with the soliton in the primary mode family, which is understood as the origin of the soliton breathing dynamics. Such an inter-mode breather soliton is distinct from the intrinsic breather soliton, as it is observed to exist in a detuning range where the regular LLE predicts stationary cavity solitons to be formed. We demonstrated the inter-mode breather soliton in two microresonator platforms, i.e. the ${\rm MgF_2}$ crystalline resonator and the chip-based ${\rm Si_3N_4}$ microresonators, and performed simulations that qualitatively confirm our experimental observations. Our results provide an new insight in the dynamics of cavity dissipative solitons by revealing an energy exchange regime that can generally exist in multi-mode Kerr cavities. Equally important, they are highly relevant to applications as inter-mode breather soliton implies a new panel of soliton instability, which is detrimental to e.g. the low-noise microwave generation.
Moreover, we can experimentally further confirm that inter-mode interactions can induce the soliton decay as well as the soliton switching. The abrupt power jump as well as the breathing within the stationary soliton existence range can perturb states with multiple DKS, leading to a decrease in the number of soliton (i.e. the soliton switching).

\section{Acknowledgments}
\begin{acknowledgments}
This publication was supported by Contract W31P4Q-14-C-0050 from the Defense Advanced Research Projects Agency (DARPA), Defense Sciences Office (DSO);
by the Air Force Office of Scientific Research, Air Force Material Command, USAF under Award No. FA9550-15-1-0099;
by funding from the European Union’s Horizon 2020;
and by the Swiss National Science Foundation under grant agreement No. 161573;
H.G. acknowledges the support from research and innovation programme under Marie Sklodowska-Curie IF grant agreement No. 709249.
E.L. and M.K. acknowledge the support from  the European Space Technology Centre with ESA Contract No. 4000116145/16/NL/MH/GM and No. 4000118777/16/NL/GM.
M.K. acknowledges the support from the European Union’s FP7 programme under under Marie Sklodowska-Curie Initial Training Network grant agreement No. 607493.
MG acknowledges support from the Hasler foundation and support from the ``EPFL Fellows'' fellowship program co-funded by Marie Curie, FP7 Grant agreement no. 291771.
${\rm Si_3N_4}$ samples were fabricated and grown in the Center of MicroNanoTechnology (CMi) at EPFL.
\end{acknowledgments}


\begin{thebibliography}{45}%
\makeatletter
\providecommand \@ifxundefined [1]{%
 \@ifx{#1\undefined}
}%
\providecommand \@ifnum [1]{%
 \ifnum #1\expandafter \@firstoftwo
 \else \expandafter \@secondoftwo
 \fi
}%
\providecommand \@ifx [1]{%
 \ifx #1\expandafter \@firstoftwo
 \else \expandafter \@secondoftwo
 \fi
}%
\providecommand \natexlab [1]{#1}%
\providecommand \enquote  [1]{``#1''}%
\providecommand \bibnamefont  [1]{#1}%
\providecommand \bibfnamefont [1]{#1}%
\providecommand \citenamefont [1]{#1}%
\providecommand \href@noop [0]{\@secondoftwo}%
\providecommand \href [0]{\begingroup \@sanitize@url \@href}%
\providecommand \@href[1]{\@@startlink{#1}\@@href}%
\providecommand \@@href[1]{\endgroup#1\@@endlink}%
\providecommand \@sanitize@url [0]{\catcode `\\12\catcode `\$12\catcode
  `\&12\catcode `\#12\catcode `\^12\catcode `\_12\catcode `\%12\relax}%
\providecommand \@@startlink[1]{}%
\providecommand \@@endlink[0]{}%
\providecommand \url  [0]{\begingroup\@sanitize@url \@url }%
\providecommand \@url [1]{\endgroup\@href {#1}{\urlprefix }}%
\providecommand \urlprefix  [0]{URL }%
\providecommand \Eprint [0]{\href }%
\providecommand \doibase [0]{http://dx.doi.org/}%
\providecommand \selectlanguage [0]{\@gobble}%
\providecommand \bibinfo  [0]{\@secondoftwo}%
\providecommand \bibfield  [0]{\@secondoftwo}%
\providecommand \translation [1]{[#1]}%
\providecommand \BibitemOpen [0]{}%
\providecommand \bibitemStop [0]{}%
\providecommand \bibitemNoStop [0]{.\EOS\space}%
\providecommand \EOS [0]{\spacefactor3000\relax}%
\providecommand \BibitemShut  [1]{\csname bibitem#1\endcsname}%
\let\auto@bib@innerbib\@empty
\bibitem [{\citenamefont {Herr}\ \emph
  {et~al.}(2014{\natexlab{a}})\citenamefont {Herr}, \citenamefont {Brasch},
  \citenamefont {Jost}, \citenamefont {Wang}, \citenamefont {Kondratiev},
  \citenamefont {Gorodetsky},\ and\ \citenamefont
  {Kippenberg}}]{herr2014temporal}%
  \BibitemOpen
  \bibfield  {author} {\bibinfo {author} {\bibfnamefont {T.}~\bibnamefont
  {Herr}}, \bibinfo {author} {\bibfnamefont {V.}~\bibnamefont {Brasch}},
  \bibinfo {author} {\bibfnamefont {J.~D.}\ \bibnamefont {Jost}}, \bibinfo
  {author} {\bibfnamefont {C.~Y.}\ \bibnamefont {Wang}}, \bibinfo {author}
  {\bibfnamefont {N.~M.}\ \bibnamefont {Kondratiev}}, \bibinfo {author}
  {\bibfnamefont {M.~L.}\ \bibnamefont {Gorodetsky}}, \ and\ \bibinfo {author}
  {\bibfnamefont {T.~J.}\ \bibnamefont {Kippenberg}},\ }\href {\doibase
  10.1038/nphoton.2013.343} {\bibfield  {journal} {\bibinfo  {journal} {Nature
  Photon.}\ }\textbf {\bibinfo {volume} {8}},\ \bibinfo {pages} {145} (\bibinfo
  {year} {2014}{\natexlab{a}})}\BibitemShut {NoStop}%
\bibitem [{\citenamefont {Brasch}\ \emph {et~al.}(2016)\citenamefont {Brasch},
  \citenamefont {Geiselmann}, \citenamefont {Herr}, \citenamefont {Lihachev},
  \citenamefont {Pfeiffer}, \citenamefont {Gorodetsky},\ and\ \citenamefont
  {Kippenberg}}]{brasch2016photonic}%
  \BibitemOpen
  \bibfield  {author} {\bibinfo {author} {\bibfnamefont {V.}~\bibnamefont
  {Brasch}}, \bibinfo {author} {\bibfnamefont {M.}~\bibnamefont {Geiselmann}},
  \bibinfo {author} {\bibfnamefont {T.}~\bibnamefont {Herr}}, \bibinfo {author}
  {\bibfnamefont {G.}~\bibnamefont {Lihachev}}, \bibinfo {author}
  {\bibfnamefont {M.~H.~P.}\ \bibnamefont {Pfeiffer}}, \bibinfo {author}
  {\bibfnamefont {M.~L.}\ \bibnamefont {Gorodetsky}}, \ and\ \bibinfo {author}
  {\bibfnamefont {T.~J.}\ \bibnamefont {Kippenberg}},\ }\href {\doibase
  10.1126/science.aad4811} {\bibfield  {journal} {\bibinfo  {journal}
  {Science}\ }\textbf {\bibinfo {volume} {351}},\ \bibinfo {pages} {357}
  (\bibinfo {year} {2016})}\BibitemShut {NoStop}%
\bibitem [{\citenamefont {Yi}\ \emph {et~al.}(2015)\citenamefont {Yi},
  \citenamefont {Yang}, \citenamefont {Yang}, \citenamefont {Suh},\ and\
  \citenamefont {Vahala}}]{yi2015soliton}%
  \BibitemOpen
  \bibfield  {author} {\bibinfo {author} {\bibfnamefont {X.}~\bibnamefont
  {Yi}}, \bibinfo {author} {\bibfnamefont {Q.-F.}\ \bibnamefont {Yang}},
  \bibinfo {author} {\bibfnamefont {K.~Y.}\ \bibnamefont {Yang}}, \bibinfo
  {author} {\bibfnamefont {M.-G.}\ \bibnamefont {Suh}}, \ and\ \bibinfo
  {author} {\bibfnamefont {K.}~\bibnamefont {Vahala}},\ }\href {\doibase
  10.1364/OPTICA.2.001078} {\bibfield  {journal} {\bibinfo  {journal} {Optica}\
  }\textbf {\bibinfo {volume} {2}},\ \bibinfo {pages} {1078} (\bibinfo {year}
  {2015})}\BibitemShut {NoStop}%
\bibitem [{\citenamefont {Liang}\ \emph {et~al.}(2015)\citenamefont {Liang},
  \citenamefont {Eliyahu}, \citenamefont {Ilchenko}, \citenamefont
  {Savchenkov}, \citenamefont {Matsko}, \citenamefont {Seidel},\ and\
  \citenamefont {Maleki}}]{liang2015high}%
  \BibitemOpen
  \bibfield  {author} {\bibinfo {author} {\bibfnamefont {W.}~\bibnamefont
  {Liang}}, \bibinfo {author} {\bibfnamefont {D.}~\bibnamefont {Eliyahu}},
  \bibinfo {author} {\bibfnamefont {V.}~\bibnamefont {Ilchenko}}, \bibinfo
  {author} {\bibfnamefont {A.}~\bibnamefont {Savchenkov}}, \bibinfo {author}
  {\bibfnamefont {A.}~\bibnamefont {Matsko}}, \bibinfo {author} {\bibfnamefont
  {D.}~\bibnamefont {Seidel}}, \ and\ \bibinfo {author} {\bibfnamefont
  {L.}~\bibnamefont {Maleki}},\ }\href {\doibase 10.1038/ncomms8957} {\bibfield
   {journal} {\bibinfo  {journal} {Nature commun.}\ }\textbf {\bibinfo {volume}
  {6}} (\bibinfo {year} {2015}),\ 10.1038/ncomms8957}\BibitemShut {NoStop}%
\bibitem [{\citenamefont {Wang}\ \emph {et~al.}(2016)\citenamefont {Wang},
  \citenamefont {Jaramillo-Villegas}, \citenamefont {Xuan}, \citenamefont
  {Xue}, \citenamefont {Bao}, \citenamefont {Leaird}, \citenamefont {Qi},\ and\
  \citenamefont {Weiner}}]{wang2016intracavity}%
  \BibitemOpen
  \bibfield  {author} {\bibinfo {author} {\bibfnamefont {P.-H.}\ \bibnamefont
  {Wang}}, \bibinfo {author} {\bibfnamefont {J.~A.}\ \bibnamefont
  {Jaramillo-Villegas}}, \bibinfo {author} {\bibfnamefont {Y.}~\bibnamefont
  {Xuan}}, \bibinfo {author} {\bibfnamefont {X.}~\bibnamefont {Xue}}, \bibinfo
  {author} {\bibfnamefont {C.}~\bibnamefont {Bao}}, \bibinfo {author}
  {\bibfnamefont {D.~E.}\ \bibnamefont {Leaird}}, \bibinfo {author}
  {\bibfnamefont {M.}~\bibnamefont {Qi}}, \ and\ \bibinfo {author}
  {\bibfnamefont {A.~M.}\ \bibnamefont {Weiner}},\ }\href {\doibase
  10.1364/OE.24.010890} {\bibfield  {journal} {\bibinfo  {journal} {Optics
  express}\ }\textbf {\bibinfo {volume} {24}},\ \bibinfo {pages} {10890}
  (\bibinfo {year} {2016})}\BibitemShut {NoStop}%
\bibitem [{\citenamefont {Joshi}\ \emph {et~al.}(2016)\citenamefont {Joshi},
  \citenamefont {Jang}, \citenamefont {Luke}, \citenamefont {Ji}, \citenamefont
  {Miller}, \citenamefont {Klenner}, \citenamefont {Okawachi}, \citenamefont
  {Lipson},\ and\ \citenamefont {Gaeta}}]{joshi2016thermally}%
  \BibitemOpen
  \bibfield  {author} {\bibinfo {author} {\bibfnamefont {C.}~\bibnamefont
  {Joshi}}, \bibinfo {author} {\bibfnamefont {J.~K.}\ \bibnamefont {Jang}},
  \bibinfo {author} {\bibfnamefont {K.}~\bibnamefont {Luke}}, \bibinfo {author}
  {\bibfnamefont {X.}~\bibnamefont {Ji}}, \bibinfo {author} {\bibfnamefont
  {S.~A.}\ \bibnamefont {Miller}}, \bibinfo {author} {\bibfnamefont
  {A.}~\bibnamefont {Klenner}}, \bibinfo {author} {\bibfnamefont
  {Y.}~\bibnamefont {Okawachi}}, \bibinfo {author} {\bibfnamefont
  {M.}~\bibnamefont {Lipson}}, \ and\ \bibinfo {author} {\bibfnamefont {A.~L.}\
  \bibnamefont {Gaeta}},\ }\href {\doibase 10.1364/OL.41.002565} {\bibfield
  {journal} {\bibinfo  {journal} {Optics letters}\ }\textbf {\bibinfo {volume}
  {41}},\ \bibinfo {pages} {2565} (\bibinfo {year} {2016})}\BibitemShut
  {NoStop}%
\bibitem [{\citenamefont {Marin-Palomo}\ \emph {et~al.}(2016)\citenamefont
  {Marin-Palomo}, \citenamefont {Kemal}, \citenamefont {Karpov}, \citenamefont
  {Kordts}, \citenamefont {Pfeifle}, \citenamefont {Pfeiffer}, \citenamefont
  {Trocha}, \citenamefont {Wolf}, \citenamefont {Brasch}, \citenamefont
  {Rosenberger}, \citenamefont {Vijayan}, \citenamefont {Freude}, \citenamefont
  {Kippenberg},\ and\ \citenamefont {Koos}}]{marin2016microresonator}%
  \BibitemOpen
  \bibfield  {author} {\bibinfo {author} {\bibfnamefont {P.}~\bibnamefont
  {Marin-Palomo}}, \bibinfo {author} {\bibfnamefont {J.~N.}\ \bibnamefont
  {Kemal}}, \bibinfo {author} {\bibfnamefont {M.}~\bibnamefont {Karpov}},
  \bibinfo {author} {\bibfnamefont {A.}~\bibnamefont {Kordts}}, \bibinfo
  {author} {\bibfnamefont {J.}~\bibnamefont {Pfeifle}}, \bibinfo {author}
  {\bibfnamefont {M.~H.~P.}\ \bibnamefont {Pfeiffer}}, \bibinfo {author}
  {\bibfnamefont {P.}~\bibnamefont {Trocha}}, \bibinfo {author} {\bibfnamefont
  {S.}~\bibnamefont {Wolf}}, \bibinfo {author} {\bibfnamefont {V.}~\bibnamefont
  {Brasch}}, \bibinfo {author} {\bibfnamefont {R.}~\bibnamefont {Rosenberger}},
  \bibinfo {author} {\bibfnamefont {K.}~\bibnamefont {Vijayan}}, \bibinfo
  {author} {\bibfnamefont {W.}~\bibnamefont {Freude}}, \bibinfo {author}
  {\bibfnamefont {T.~J.}\ \bibnamefont {Kippenberg}}, \ and\ \bibinfo {author}
  {\bibfnamefont {C.}~\bibnamefont {Koos}},\ }\href@noop {} {\bibfield
  {journal} {\bibinfo  {journal} {arXiv: 1610.01484}\ } (\bibinfo {year}
  {2016})}\BibitemShut {NoStop}%
\bibitem [{\citenamefont {Suh}\ \emph {et~al.}(2016)\citenamefont {Suh},
  \citenamefont {Yang}, \citenamefont {Yang}, \citenamefont {Yi},\ and\
  \citenamefont {Vahala}}]{suh2016microresonator}%
  \BibitemOpen
  \bibfield  {author} {\bibinfo {author} {\bibfnamefont {M.-G.}\ \bibnamefont
  {Suh}}, \bibinfo {author} {\bibfnamefont {Q.-F.}\ \bibnamefont {Yang}},
  \bibinfo {author} {\bibfnamefont {K.-Y.}\ \bibnamefont {Yang}}, \bibinfo
  {author} {\bibfnamefont {X.}~\bibnamefont {Yi}}, \ and\ \bibinfo {author}
  {\bibfnamefont {K.~J.}\ \bibnamefont {Vahala}},\ }\href {\doibase
  10.1126/science.aah6516} {\bibfield  {journal} {\bibinfo  {journal}
  {Science}\ }\textbf {\bibinfo {volume} {354}},\ \bibinfo {pages} {600}
  (\bibinfo {year} {2016})}\BibitemShut {NoStop}%
\bibitem [{\citenamefont {Dutt}\ \emph {et~al.}(2016)\citenamefont {Dutt},
  \citenamefont {Joshi}, \citenamefont {Ji}, \citenamefont {Cardenas},
  \citenamefont {Okawachi}, \citenamefont {Luke}, \citenamefont {Gaeta},\ and\
  \citenamefont {Lipson}}]{dutt2016chip}%
  \BibitemOpen
  \bibfield  {author} {\bibinfo {author} {\bibfnamefont {A.}~\bibnamefont
  {Dutt}}, \bibinfo {author} {\bibfnamefont {C.}~\bibnamefont {Joshi}},
  \bibinfo {author} {\bibfnamefont {X.}~\bibnamefont {Ji}}, \bibinfo {author}
  {\bibfnamefont {J.}~\bibnamefont {Cardenas}}, \bibinfo {author}
  {\bibfnamefont {Y.}~\bibnamefont {Okawachi}}, \bibinfo {author}
  {\bibfnamefont {K.}~\bibnamefont {Luke}}, \bibinfo {author} {\bibfnamefont
  {A.~L.}\ \bibnamefont {Gaeta}}, \ and\ \bibinfo {author} {\bibfnamefont
  {M.}~\bibnamefont {Lipson}},\ }\href@noop {} {\bibfield  {journal} {\bibinfo
  {journal} {arXiv:1611.07673}\ } (\bibinfo {year} {2016})}\BibitemShut
  {NoStop}%
\bibitem [{\citenamefont {Jost}\ \emph {et~al.}(2015)\citenamefont {Jost},
  \citenamefont {Herr}, \citenamefont {Lecaplain}, \citenamefont {Brasch},
  \citenamefont {Pfeiffer},\ and\ \citenamefont
  {Kippenberg}}]{jost2015counting}%
  \BibitemOpen
  \bibfield  {author} {\bibinfo {author} {\bibfnamefont {J.~D.}\ \bibnamefont
  {Jost}}, \bibinfo {author} {\bibfnamefont {T.}~\bibnamefont {Herr}}, \bibinfo
  {author} {\bibfnamefont {C.}~\bibnamefont {Lecaplain}}, \bibinfo {author}
  {\bibfnamefont {V.}~\bibnamefont {Brasch}}, \bibinfo {author} {\bibfnamefont
  {M.~H.~P.}\ \bibnamefont {Pfeiffer}}, \ and\ \bibinfo {author} {\bibfnamefont
  {T.~J.}\ \bibnamefont {Kippenberg}},\ }\href {\doibase
  10.1364/OPTICA.2.000706} {\bibfield  {journal} {\bibinfo  {journal} {Optica}\
  }\textbf {\bibinfo {volume} {2}},\ \bibinfo {pages} {706} (\bibinfo {year}
  {2015})}\BibitemShut {NoStop}%
\bibitem [{\citenamefont {Brasch}\ \emph {et~al.}(2017)\citenamefont {Brasch},
  \citenamefont {Lucas}, \citenamefont {Jost}, \citenamefont {Geiselmann},\
  and\ \citenamefont {Kippenberg}}]{brasch2017self}%
  \BibitemOpen
  \bibfield  {author} {\bibinfo {author} {\bibfnamefont {V.}~\bibnamefont
  {Brasch}}, \bibinfo {author} {\bibfnamefont {E.}~\bibnamefont {Lucas}},
  \bibinfo {author} {\bibfnamefont {J.~D.}\ \bibnamefont {Jost}}, \bibinfo
  {author} {\bibfnamefont {M.}~\bibnamefont {Geiselmann}}, \ and\ \bibinfo
  {author} {\bibfnamefont {T.~J.}\ \bibnamefont {Kippenberg}},\ }\href
  {\doibase 10.1038/lsa.2016.202} {\bibfield  {journal} {\bibinfo  {journal}
  {Light Sci. Appl.}\ }\textbf {\bibinfo {volume} {6}},\ \bibinfo {pages}
  {e16202} (\bibinfo {year} {2017})}\BibitemShut {NoStop}%
\bibitem [{\citenamefont {Akhmediev}\ and\ \citenamefont
  {Ankiewicz}(2008)}]{akhmediev2008dissipative}%
  \BibitemOpen
  \bibfield  {author} {\bibinfo {author} {\bibfnamefont {N.}~\bibnamefont
  {Akhmediev}}\ and\ \bibinfo {author} {\bibfnamefont {A.}~\bibnamefont
  {Ankiewicz}},\ }\href {https://books.google.de/books?id=uX1j5bmLVVAC} {\emph
  {\bibinfo {title} {{Dissipative Solitons: From Optics to Biology and
  Medicine}}}},\ Lecture Notes in Physics\ (\bibinfo  {publisher} {Springer},\
  \bibinfo {year} {2008})\BibitemShut {NoStop}%
\bibitem [{\citenamefont {Bao}\ \emph {et~al.}(2015)\citenamefont {Bao},
  \citenamefont {Chang}, \citenamefont {Yang}, \citenamefont {Akhmediev},\ and\
  \citenamefont {Cundiff}}]{bao2015observation}%
  \BibitemOpen
  \bibfield  {author} {\bibinfo {author} {\bibfnamefont {C.}~\bibnamefont
  {Bao}}, \bibinfo {author} {\bibfnamefont {W.}~\bibnamefont {Chang}}, \bibinfo
  {author} {\bibfnamefont {C.}~\bibnamefont {Yang}}, \bibinfo {author}
  {\bibfnamefont {N.}~\bibnamefont {Akhmediev}}, \ and\ \bibinfo {author}
  {\bibfnamefont {S.~T.}\ \bibnamefont {Cundiff}},\ }\href {\doibase
  10.1103/PhysRevLett.115.253903} {\bibfield  {journal} {\bibinfo  {journal}
  {Phys. Rev. Lett.}\ }\textbf {\bibinfo {volume} {115}},\ \bibinfo {pages}
  {253903} (\bibinfo {year} {2015})}\BibitemShut {NoStop}%
\bibitem [{\citenamefont {Herink}\ \emph {et~al.}(2017)\citenamefont {Herink},
  \citenamefont {Kurtz}, \citenamefont {Jalali}, \citenamefont {Solli},\ and\
  \citenamefont {Ropers}}]{herink2017real}%
  \BibitemOpen
  \bibfield  {author} {\bibinfo {author} {\bibfnamefont {G.}~\bibnamefont
  {Herink}}, \bibinfo {author} {\bibfnamefont {F.}~\bibnamefont {Kurtz}},
  \bibinfo {author} {\bibfnamefont {B.}~\bibnamefont {Jalali}}, \bibinfo
  {author} {\bibfnamefont {D.}~\bibnamefont {Solli}}, \ and\ \bibinfo {author}
  {\bibfnamefont {C.}~\bibnamefont {Ropers}},\ }\href {\doibase
  10.1126/science.aal5326} {\bibfield  {journal} {\bibinfo  {journal}
  {Science}\ }\textbf {\bibinfo {volume} {356}},\ \bibinfo {pages} {50}
  (\bibinfo {year} {2017})}\BibitemShut {NoStop}%
\bibitem [{\citenamefont {Leo}\ \emph {et~al.}(2013)\citenamefont {Leo},
  \citenamefont {Gelens}, \citenamefont {Emplit}, \citenamefont {Haelterman},\
  and\ \citenamefont {Coen}}]{leo2013dynamics}%
  \BibitemOpen
  \bibfield  {author} {\bibinfo {author} {\bibfnamefont {F.}~\bibnamefont
  {Leo}}, \bibinfo {author} {\bibfnamefont {L.}~\bibnamefont {Gelens}},
  \bibinfo {author} {\bibfnamefont {P.}~\bibnamefont {Emplit}}, \bibinfo
  {author} {\bibfnamefont {M.}~\bibnamefont {Haelterman}}, \ and\ \bibinfo
  {author} {\bibfnamefont {S.}~\bibnamefont {Coen}},\ }\href {\doibase
  10.1364/OE.21.009180} {\bibfield  {journal} {\bibinfo  {journal} {Opt.
  Express}\ }\textbf {\bibinfo {volume} {21}},\ \bibinfo {pages} {9180}
  (\bibinfo {year} {2013})}\BibitemShut {NoStop}%
\bibitem [{\citenamefont {Bao}\ \emph {et~al.}(2016)\citenamefont {Bao},
  \citenamefont {Jaramillo-Villegas}, \citenamefont {Xuan}, \citenamefont
  {Leaird}, \citenamefont {Qi},\ and\ \citenamefont
  {Weiner}}]{bao2016observation}%
  \BibitemOpen
  \bibfield  {author} {\bibinfo {author} {\bibfnamefont {C.}~\bibnamefont
  {Bao}}, \bibinfo {author} {\bibfnamefont {J.~A.}\ \bibnamefont
  {Jaramillo-Villegas}}, \bibinfo {author} {\bibfnamefont {Y.}~\bibnamefont
  {Xuan}}, \bibinfo {author} {\bibfnamefont {D.~E.}\ \bibnamefont {Leaird}},
  \bibinfo {author} {\bibfnamefont {M.}~\bibnamefont {Qi}}, \ and\ \bibinfo
  {author} {\bibfnamefont {A.~M.}\ \bibnamefont {Weiner}},\ }\href {\doibase
  10.1103/PhysRevLett.117.163901} {\bibfield  {journal} {\bibinfo  {journal}
  {Phys. Rev. Lett.}\ }\textbf {\bibinfo {volume} {117}},\ \bibinfo {pages}
  {163901} (\bibinfo {year} {2016})}\BibitemShut {NoStop}%
\bibitem [{\citenamefont {Yu}\ \emph {et~al.}(2017)\citenamefont {Yu},
  \citenamefont {Jang}, \citenamefont {Okawachi}, \citenamefont {Griffith},
  \citenamefont {Luke}, \citenamefont {Miller}, \citenamefont {Ji},
  \citenamefont {Lipson},\ and\ \citenamefont {Gaeta}}]{yu2016breather}%
  \BibitemOpen
  \bibfield  {author} {\bibinfo {author} {\bibfnamefont {M.}~\bibnamefont
  {Yu}}, \bibinfo {author} {\bibfnamefont {J.~K.}\ \bibnamefont {Jang}},
  \bibinfo {author} {\bibfnamefont {Y.}~\bibnamefont {Okawachi}}, \bibinfo
  {author} {\bibfnamefont {A.~G.}\ \bibnamefont {Griffith}}, \bibinfo {author}
  {\bibfnamefont {K.}~\bibnamefont {Luke}}, \bibinfo {author} {\bibfnamefont
  {S.~A.}\ \bibnamefont {Miller}}, \bibinfo {author} {\bibfnamefont
  {X.}~\bibnamefont {Ji}}, \bibinfo {author} {\bibfnamefont {M.}~\bibnamefont
  {Lipson}}, \ and\ \bibinfo {author} {\bibfnamefont {A.~L.}\ \bibnamefont
  {Gaeta}},\ }\href {\doibase 10.1038/ncomms14569} {\bibfield  {journal}
  {\bibinfo  {journal} {Nature Commun.}\ }\textbf {\bibinfo {volume} {8}}
  (\bibinfo {year} {2017}),\ 10.1038/ncomms14569}\BibitemShut {NoStop}%
\bibitem [{\citenamefont {Lucas}\ \emph {et~al.}(2016)\citenamefont {Lucas},
  \citenamefont {Karpov}, \citenamefont {Guo}, \citenamefont {Gorodetsky},\
  and\ \citenamefont {Kippenberg}}]{lucas2016breathing}%
  \BibitemOpen
  \bibfield  {author} {\bibinfo {author} {\bibfnamefont {E.}~\bibnamefont
  {Lucas}}, \bibinfo {author} {\bibfnamefont {M.}~\bibnamefont {Karpov}},
  \bibinfo {author} {\bibfnamefont {H.}~\bibnamefont {Guo}}, \bibinfo {author}
  {\bibfnamefont {M.~L.}\ \bibnamefont {Gorodetsky}}, \ and\ \bibinfo {author}
  {\bibfnamefont {T.~J.}\ \bibnamefont {Kippenberg}},\ }\href@noop {}
  {\bibfield  {journal} {\bibinfo  {journal} {arXiv: 1611.06567}\ } (\bibinfo
  {year} {2016})}\BibitemShut {NoStop}%
\bibitem [{\citenamefont {Barashenkov}\ and\ \citenamefont
  {Smirnov}(1996)}]{barashenkov1996existence}%
  \BibitemOpen
  \bibfield  {author} {\bibinfo {author} {\bibfnamefont {I.~V.}\ \bibnamefont
  {Barashenkov}}\ and\ \bibinfo {author} {\bibfnamefont {Y.~S.}\ \bibnamefont
  {Smirnov}},\ }\href {\doibase 10.1103/PhysRevE.54.5707} {\bibfield  {journal}
  {\bibinfo  {journal} {Phys. Rev. E}\ }\textbf {\bibinfo {volume} {54}},\
  \bibinfo {pages} {5707} (\bibinfo {year} {1996})}\BibitemShut {NoStop}%
\bibitem [{\citenamefont {Parra-Rivas}\ \emph {et~al.}(2014)\citenamefont
  {Parra-Rivas}, \citenamefont {Gomila}, \citenamefont {Mat{\'\i}as},
  \citenamefont {Coen},\ and\ \citenamefont {Gelens}}]{parra2014dynamics}%
  \BibitemOpen
  \bibfield  {author} {\bibinfo {author} {\bibfnamefont {P.}~\bibnamefont
  {Parra-Rivas}}, \bibinfo {author} {\bibfnamefont {D.}~\bibnamefont {Gomila}},
  \bibinfo {author} {\bibfnamefont {M.~A.}\ \bibnamefont {Mat{\'\i}as}},
  \bibinfo {author} {\bibfnamefont {S.}~\bibnamefont {Coen}}, \ and\ \bibinfo
  {author} {\bibfnamefont {L.}~\bibnamefont {Gelens}},\ }\href {\doibase
  10.1103/PhysRevA.89.043813} {\bibfield  {journal} {\bibinfo  {journal} {Phys.
  Rev. A}\ }\textbf {\bibinfo {volume} {89}},\ \bibinfo {pages} {043813}
  (\bibinfo {year} {2014})}\BibitemShut {NoStop}%
\bibitem [{\citenamefont {Godey}\ \emph {et~al.}(2014)\citenamefont {Godey},
  \citenamefont {Balakireva}, \citenamefont {Coillet},\ and\ \citenamefont
  {Chembo}}]{godey2014stability}%
  \BibitemOpen
  \bibfield  {author} {\bibinfo {author} {\bibfnamefont {C.}~\bibnamefont
  {Godey}}, \bibinfo {author} {\bibfnamefont {I.~V.}\ \bibnamefont
  {Balakireva}}, \bibinfo {author} {\bibfnamefont {A.}~\bibnamefont {Coillet}},
  \ and\ \bibinfo {author} {\bibfnamefont {Y.~K.}\ \bibnamefont {Chembo}},\
  }\href {\doibase 10.1103/PhysRevA.89.063814} {\bibfield  {journal} {\bibinfo
  {journal} {Phys. Rev. A}\ }\textbf {\bibinfo {volume} {89}},\ \bibinfo
  {pages} {063814} (\bibinfo {year} {2014})}\BibitemShut {NoStop}%
\bibitem [{\citenamefont {Lugiato}\ and\ \citenamefont
  {Lefever}(1987)}]{lugiato1987spatial}%
  \BibitemOpen
  \bibfield  {author} {\bibinfo {author} {\bibfnamefont {L.~A.}\ \bibnamefont
  {Lugiato}}\ and\ \bibinfo {author} {\bibfnamefont {R.}~\bibnamefont
  {Lefever}},\ }\href {\doibase 10.1103/PhysRevLett.58.2209} {\bibfield
  {journal} {\bibinfo  {journal} {Phys. Rev. Lett.}\ }\textbf {\bibinfo
  {volume} {58}},\ \bibinfo {pages} {2209} (\bibinfo {year}
  {1987})}\BibitemShut {NoStop}%
\bibitem [{\citenamefont {Chembo}\ and\ \citenamefont
  {Yu}(2010)}]{chembo2010modal}%
  \BibitemOpen
  \bibfield  {author} {\bibinfo {author} {\bibfnamefont {Y.~K.}\ \bibnamefont
  {Chembo}}\ and\ \bibinfo {author} {\bibfnamefont {N.}~\bibnamefont {Yu}},\
  }\href {\doibase 10.1103/PhysRevA.82.033801} {\bibfield  {journal} {\bibinfo
  {journal} {Phys. Rev. A}\ }\textbf {\bibinfo {volume} {82}},\ \bibinfo
  {pages} {033801} (\bibinfo {year} {2010})}\BibitemShut {NoStop}%
\bibitem [{\citenamefont {Coen}\ \emph {et~al.}(2013)\citenamefont {Coen},
  \citenamefont {Randle}, \citenamefont {Sylvestre},\ and\ \citenamefont
  {Erkintalo}}]{coen2013modeling}%
  \BibitemOpen
  \bibfield  {author} {\bibinfo {author} {\bibfnamefont {S.}~\bibnamefont
  {Coen}}, \bibinfo {author} {\bibfnamefont {H.~G.}\ \bibnamefont {Randle}},
  \bibinfo {author} {\bibfnamefont {T.}~\bibnamefont {Sylvestre}}, \ and\
  \bibinfo {author} {\bibfnamefont {M.}~\bibnamefont {Erkintalo}},\ }\href
  {\doibase 10.1364/OL.38.000037} {\bibfield  {journal} {\bibinfo  {journal}
  {Opt. Lett.}\ }\textbf {\bibinfo {volume} {38}},\ \bibinfo {pages} {37}
  (\bibinfo {year} {2013})}\BibitemShut {NoStop}%
\bibitem [{\citenamefont {Chembo}\ and\ \citenamefont
  {Menyuk}(2013)}]{chembo2013spatio}%
  \BibitemOpen
  \bibfield  {author} {\bibinfo {author} {\bibfnamefont {Y.~K.}\ \bibnamefont
  {Chembo}}\ and\ \bibinfo {author} {\bibfnamefont {C.~R.}\ \bibnamefont
  {Menyuk}},\ }\href {\doibase 10.1103/PhysRevA.87.053852} {\bibfield
  {journal} {\bibinfo  {journal} {Phys. Rev. A}\ }\textbf {\bibinfo {volume}
  {87}},\ \bibinfo {pages} {053852} (\bibinfo {year} {2013})}\BibitemShut
  {NoStop}%
\bibitem [{\citenamefont {Lucas}\ \emph {et~al.}(2017)\citenamefont {Lucas},
  \citenamefont {Guo}, \citenamefont {Jost}, \citenamefont {Karpov},\ and\
  \citenamefont {Kippenberg}}]{lucas2016study}%
  \BibitemOpen
  \bibfield  {author} {\bibinfo {author} {\bibfnamefont {E.}~\bibnamefont
  {Lucas}}, \bibinfo {author} {\bibfnamefont {H.}~\bibnamefont {Guo}}, \bibinfo
  {author} {\bibfnamefont {J.~D.}\ \bibnamefont {Jost}}, \bibinfo {author}
  {\bibfnamefont {M.}~\bibnamefont {Karpov}}, \ and\ \bibinfo {author}
  {\bibfnamefont {T.~J.}\ \bibnamefont {Kippenberg}},\ }\href {\doibase
  10.1103/PhysRevA.95.043822} {\bibfield  {journal} {\bibinfo  {journal} {Phys.
  Rev. A}\ }\textbf {\bibinfo {volume} {95}},\ \bibinfo {pages} {043822}
  (\bibinfo {year} {2017})}\BibitemShut {NoStop}%
\bibitem [{\citenamefont {Guo}\ \emph {et~al.}(2017)\citenamefont {Guo},
  \citenamefont {Karpov}, \citenamefont {Lucas}, \citenamefont {Kordts},
  \citenamefont {Pfeiffer}, \citenamefont {Brasch}, \citenamefont {Lihachev},
  \citenamefont {Lobanov}, \citenamefont {Gorodetsky},\ and\ \citenamefont
  {Kippenberg}}]{guo2016universal}%
  \BibitemOpen
  \bibfield  {author} {\bibinfo {author} {\bibfnamefont {H.}~\bibnamefont
  {Guo}}, \bibinfo {author} {\bibfnamefont {M.}~\bibnamefont {Karpov}},
  \bibinfo {author} {\bibfnamefont {E.}~\bibnamefont {Lucas}}, \bibinfo
  {author} {\bibfnamefont {A.}~\bibnamefont {Kordts}}, \bibinfo {author}
  {\bibfnamefont {M.~H.~P.}\ \bibnamefont {Pfeiffer}}, \bibinfo {author}
  {\bibfnamefont {V.}~\bibnamefont {Brasch}}, \bibinfo {author} {\bibfnamefont
  {G.}~\bibnamefont {Lihachev}}, \bibinfo {author} {\bibfnamefont {V.~E.}\
  \bibnamefont {Lobanov}}, \bibinfo {author} {\bibfnamefont {M.~L.}\
  \bibnamefont {Gorodetsky}}, \ and\ \bibinfo {author} {\bibfnamefont {T.~J.}\
  \bibnamefont {Kippenberg}},\ }\href {\doibase 10.1038/nphys3893} {\bibfield
  {journal} {\bibinfo  {journal} {Nature Phys.}\ }\textbf {\bibinfo {volume}
  {13}},\ \bibinfo {pages} {94} (\bibinfo {year} {2017})}\BibitemShut {NoStop}%
\bibitem [{\citenamefont {Riemensberger}\ \emph {et~al.}(2012)\citenamefont
  {Riemensberger}, \citenamefont {Hartinger}, \citenamefont {Herr},
  \citenamefont {Brasch}, \citenamefont {Holzwarth},\ and\ \citenamefont
  {Kippenberg}}]{riemensberger2012dispersion}%
  \BibitemOpen
  \bibfield  {author} {\bibinfo {author} {\bibfnamefont {J.}~\bibnamefont
  {Riemensberger}}, \bibinfo {author} {\bibfnamefont {K.}~\bibnamefont
  {Hartinger}}, \bibinfo {author} {\bibfnamefont {T.}~\bibnamefont {Herr}},
  \bibinfo {author} {\bibfnamefont {V.}~\bibnamefont {Brasch}}, \bibinfo
  {author} {\bibfnamefont {R.}~\bibnamefont {Holzwarth}}, \ and\ \bibinfo
  {author} {\bibfnamefont {T.~J.}\ \bibnamefont {Kippenberg}},\ }\href
  {\doibase 10.1364/OE.20.027661} {\bibfield  {journal} {\bibinfo  {journal}
  {Opt. Express}\ }\textbf {\bibinfo {volume} {20}},\ \bibinfo {pages} {27661}
  (\bibinfo {year} {2012})}\BibitemShut {NoStop}%
\bibitem [{\citenamefont {Pfeiffer}\ \emph {et~al.}(2016)\citenamefont
  {Pfeiffer}, \citenamefont {Kordts}, \citenamefont {Brasch}, \citenamefont
  {Zervas}, \citenamefont {Geiselmann}, \citenamefont {Jost},\ and\
  \citenamefont {Kippenberg}}]{pfeiffer2016photonic}%
  \BibitemOpen
  \bibfield  {author} {\bibinfo {author} {\bibfnamefont {M.~H.~P.}\
  \bibnamefont {Pfeiffer}}, \bibinfo {author} {\bibfnamefont {A.}~\bibnamefont
  {Kordts}}, \bibinfo {author} {\bibfnamefont {V.}~\bibnamefont {Brasch}},
  \bibinfo {author} {\bibfnamefont {M.}~\bibnamefont {Zervas}}, \bibinfo
  {author} {\bibfnamefont {M.}~\bibnamefont {Geiselmann}}, \bibinfo {author}
  {\bibfnamefont {J.~D.}\ \bibnamefont {Jost}}, \ and\ \bibinfo {author}
  {\bibfnamefont {T.~J.}\ \bibnamefont {Kippenberg}},\ }\href {\doibase
  10.1364/OPTICA.3.000020} {\bibfield  {journal} {\bibinfo  {journal} {Optica}\
  }\textbf {\bibinfo {volume} {3}},\ \bibinfo {pages} {20} (\bibinfo {year}
  {2016})}\BibitemShut {NoStop}%
\bibitem [{\citenamefont {Carmon}\ \emph {et~al.}(2008)\citenamefont {Carmon},
  \citenamefont {Schwefel}, \citenamefont {Yang}, \citenamefont {Oxborrow},
  \citenamefont {Stone},\ and\ \citenamefont {Vahala}}]{carmon2008static}%
  \BibitemOpen
  \bibfield  {author} {\bibinfo {author} {\bibfnamefont {T.}~\bibnamefont
  {Carmon}}, \bibinfo {author} {\bibfnamefont {H.~G.~L.}\ \bibnamefont
  {Schwefel}}, \bibinfo {author} {\bibfnamefont {L.}~\bibnamefont {Yang}},
  \bibinfo {author} {\bibfnamefont {M.}~\bibnamefont {Oxborrow}}, \bibinfo
  {author} {\bibfnamefont {A.~D.}\ \bibnamefont {Stone}}, \ and\ \bibinfo
  {author} {\bibfnamefont {K.~J.}\ \bibnamefont {Vahala}},\ }\href {\doibase
  10.1103/PhysRevLett.100.103905} {\bibfield  {journal} {\bibinfo  {journal}
  {Phys. Rev. Lett.}\ }\textbf {\bibinfo {volume} {100}},\ \bibinfo {pages}
  {103905} (\bibinfo {year} {2008})}\BibitemShut {NoStop}%
\bibitem [{\citenamefont {Herr}\ \emph
  {et~al.}(2014{\natexlab{b}})\citenamefont {Herr}, \citenamefont {Brasch},
  \citenamefont {Jost}, \citenamefont {Mirgorodskiy}, \citenamefont {Lihachev},
  \citenamefont {Gorodetsky},\ and\ \citenamefont {Kippenberg}}]{herr2014mode}%
  \BibitemOpen
  \bibfield  {author} {\bibinfo {author} {\bibfnamefont {T.}~\bibnamefont
  {Herr}}, \bibinfo {author} {\bibfnamefont {V.}~\bibnamefont {Brasch}},
  \bibinfo {author} {\bibfnamefont {J.~D.}\ \bibnamefont {Jost}}, \bibinfo
  {author} {\bibfnamefont {I.}~\bibnamefont {Mirgorodskiy}}, \bibinfo {author}
  {\bibfnamefont {G.}~\bibnamefont {Lihachev}}, \bibinfo {author}
  {\bibfnamefont {M.~L.}\ \bibnamefont {Gorodetsky}}, \ and\ \bibinfo {author}
  {\bibfnamefont {T.~J.}\ \bibnamefont {Kippenberg}},\ }\href {\doibase
  10.1103/PhysRevLett.113.123901} {\bibfield  {journal} {\bibinfo  {journal}
  {Phys. Rev. Lett.}\ }\textbf {\bibinfo {volume} {113}},\ \bibinfo {pages}
  {123901} (\bibinfo {year} {2014}{\natexlab{b}})}\BibitemShut {NoStop}%
\bibitem [{\citenamefont {Kordts}\ \emph {et~al.}(2016)\citenamefont {Kordts},
  \citenamefont {Pfeiffer}, \citenamefont {Guo}, \citenamefont {Brasch},\ and\
  \citenamefont {Kippenberg}}]{kordts2016higher}%
  \BibitemOpen
  \bibfield  {author} {\bibinfo {author} {\bibfnamefont {A.}~\bibnamefont
  {Kordts}}, \bibinfo {author} {\bibfnamefont {M.~H.~P.}\ \bibnamefont
  {Pfeiffer}}, \bibinfo {author} {\bibfnamefont {H.}~\bibnamefont {Guo}},
  \bibinfo {author} {\bibfnamefont {V.}~\bibnamefont {Brasch}}, \ and\ \bibinfo
  {author} {\bibfnamefont {T.~J.}\ \bibnamefont {Kippenberg}},\ }\href
  {\doibase doi.org/10.1364/OL.41.000452} {\bibfield  {journal} {\bibinfo
  {journal} {Opt. Lett.}\ }\textbf {\bibinfo {volume} {41}},\ \bibinfo {pages}
  {452} (\bibinfo {year} {2016})}\BibitemShut {NoStop}%
\bibitem [{\citenamefont {Wabnitz}(1993)}]{wabnitz1993suppression}%
  \BibitemOpen
  \bibfield  {author} {\bibinfo {author} {\bibfnamefont {S.}~\bibnamefont
  {Wabnitz}},\ }\href {\doibase 10.1364/OL.18.000601} {\bibfield  {journal}
  {\bibinfo  {journal} {Opt. Lett.}\ }\textbf {\bibinfo {volume} {18}},\
  \bibinfo {pages} {601} (\bibinfo {year} {1993})}\BibitemShut {NoStop}%
\bibitem [{\citenamefont {Matsko}\ \emph {et~al.}(2016)\citenamefont {Matsko},
  \citenamefont {Liang}, \citenamefont {Savchenkov}, \citenamefont {Eliyahu},\
  and\ \citenamefont {Maleki}}]{matsko2016optical}%
  \BibitemOpen
  \bibfield  {author} {\bibinfo {author} {\bibfnamefont {A.~B.}\ \bibnamefont
  {Matsko}}, \bibinfo {author} {\bibfnamefont {W.}~\bibnamefont {Liang}},
  \bibinfo {author} {\bibfnamefont {A.~A.}\ \bibnamefont {Savchenkov}},
  \bibinfo {author} {\bibfnamefont {D.}~\bibnamefont {Eliyahu}}, \ and\
  \bibinfo {author} {\bibfnamefont {L.}~\bibnamefont {Maleki}},\ }\href
  {\doibase 10.1364/OL.41.002907} {\bibfield  {journal} {\bibinfo  {journal}
  {Opt. Lett.}\ }\textbf {\bibinfo {volume} {41}},\ \bibinfo {pages} {2907}
  (\bibinfo {year} {2016})}\BibitemShut {NoStop}%
\bibitem [{\citenamefont {Yang}\ \emph {et~al.}(2016)\citenamefont {Yang},
  \citenamefont {Yi}, \citenamefont {Yang},\ and\ \citenamefont
  {Vahala}}]{yang2016spatial}%
  \BibitemOpen
  \bibfield  {author} {\bibinfo {author} {\bibfnamefont {Q.-F.}\ \bibnamefont
  {Yang}}, \bibinfo {author} {\bibfnamefont {X.}~\bibnamefont {Yi}}, \bibinfo
  {author} {\bibfnamefont {K.-Y.}\ \bibnamefont {Yang}}, \ and\ \bibinfo
  {author} {\bibfnamefont {K.~J.}\ \bibnamefont {Vahala}},\ }\href {\doibase
  10.1364/OPTICA.3.001132} {\bibfield  {journal} {\bibinfo  {journal} {Optica}\
  }\textbf {\bibinfo {volume} {3}},\ \bibinfo {pages} {1132} (\bibinfo {year}
  {2016})}\BibitemShut {NoStop}%
\bibitem [{\citenamefont {Yi}\ \emph {et~al.}(2016{\natexlab{a}})\citenamefont
  {Yi}, \citenamefont {Yang}, \citenamefont {Zhang}, \citenamefont {Yang},\
  and\ \citenamefont {Vahala}}]{yi2016single}%
  \BibitemOpen
  \bibfield  {author} {\bibinfo {author} {\bibfnamefont {X.}~\bibnamefont
  {Yi}}, \bibinfo {author} {\bibfnamefont {Q.-F.}\ \bibnamefont {Yang}},
  \bibinfo {author} {\bibfnamefont {X.}~\bibnamefont {Zhang}}, \bibinfo
  {author} {\bibfnamefont {K.-Y.}\ \bibnamefont {Yang}}, \ and\ \bibinfo
  {author} {\bibfnamefont {K.~J.}\ \bibnamefont {Vahala}},\ }\href@noop {}
  {\bibfield  {journal} {\bibinfo  {journal} {arXiv: 1610.08145}\ } (\bibinfo
  {year} {2016}{\natexlab{a}})}\BibitemShut {NoStop}%
\bibitem [{\citenamefont {Xue}\ \emph {et~al.}(2015)\citenamefont {Xue},
  \citenamefont {Xuan}, \citenamefont {Wang}, \citenamefont {Liu},
  \citenamefont {Leaird}, \citenamefont {Qi},\ and\ \citenamefont
  {Weiner}}]{xue2015normal}%
  \BibitemOpen
  \bibfield  {author} {\bibinfo {author} {\bibfnamefont {X.}~\bibnamefont
  {Xue}}, \bibinfo {author} {\bibfnamefont {Y.}~\bibnamefont {Xuan}}, \bibinfo
  {author} {\bibfnamefont {P.-H.}\ \bibnamefont {Wang}}, \bibinfo {author}
  {\bibfnamefont {Y.}~\bibnamefont {Liu}}, \bibinfo {author} {\bibfnamefont
  {D.~E.}\ \bibnamefont {Leaird}}, \bibinfo {author} {\bibfnamefont
  {M.}~\bibnamefont {Qi}}, \ and\ \bibinfo {author} {\bibfnamefont {A.~M.}\
  \bibnamefont {Weiner}},\ }\href {\doibase 10.1002/lpor.201500107} {\bibfield
  {journal} {\bibinfo  {journal} {Laser Photon. Rev.}\ }\textbf {\bibinfo
  {volume} {9}},\ \bibinfo {pages} {L23} (\bibinfo {year} {2015})}\BibitemShut
  {NoStop}%
\bibitem [{\citenamefont {Miller}\ \emph {et~al.}(2015)\citenamefont {Miller},
  \citenamefont {Okawachi}, \citenamefont {Ramelow}, \citenamefont {Luke},
  \citenamefont {Dutt}, \citenamefont {Farsi}, \citenamefont {Gaeta},\ and\
  \citenamefont {Lipson}}]{miller2015tunable}%
  \BibitemOpen
  \bibfield  {author} {\bibinfo {author} {\bibfnamefont {S.~A.}\ \bibnamefont
  {Miller}}, \bibinfo {author} {\bibfnamefont {Y.}~\bibnamefont {Okawachi}},
  \bibinfo {author} {\bibfnamefont {S.}~\bibnamefont {Ramelow}}, \bibinfo
  {author} {\bibfnamefont {K.}~\bibnamefont {Luke}}, \bibinfo {author}
  {\bibfnamefont {A.}~\bibnamefont {Dutt}}, \bibinfo {author} {\bibfnamefont
  {A.}~\bibnamefont {Farsi}}, \bibinfo {author} {\bibfnamefont {A.~L.}\
  \bibnamefont {Gaeta}}, \ and\ \bibinfo {author} {\bibfnamefont
  {M.}~\bibnamefont {Lipson}},\ }\href {\doibase 10.1364/OE.23.021527}
  {\bibfield  {journal} {\bibinfo  {journal} {Opt. Express}\ }\textbf {\bibinfo
  {volume} {23}},\ \bibinfo {pages} {21527} (\bibinfo {year}
  {2015})}\BibitemShut {NoStop}%
\bibitem [{\citenamefont {Soltani}\ \emph {et~al.}(2016)\citenamefont
  {Soltani}, \citenamefont {Matsko},\ and\ \citenamefont
  {Maleki}}]{soltani2016enabling}%
  \BibitemOpen
  \bibfield  {author} {\bibinfo {author} {\bibfnamefont {M.}~\bibnamefont
  {Soltani}}, \bibinfo {author} {\bibfnamefont {A.~B.}\ \bibnamefont {Matsko}},
  \ and\ \bibinfo {author} {\bibfnamefont {L.}~\bibnamefont {Maleki}},\ }\href
  {\doibase 10.1002/lpor.201500226} {\bibfield  {journal} {\bibinfo  {journal}
  {Laser Photon. Rev.}\ }\textbf {\bibinfo {volume} {10}},\ \bibinfo {pages}
  {158} (\bibinfo {year} {2016})}\BibitemShut {NoStop}%
\bibitem [{\citenamefont {Karpov}\ \emph {et~al.}(2016)\citenamefont {Karpov},
  \citenamefont {Guo}, \citenamefont {Kordts}, \citenamefont {Brasch},
  \citenamefont {Pfeiffer}, \citenamefont {Zervas}, \citenamefont
  {Geiselmann},\ and\ \citenamefont {Kippenberg}}]{karpov2016raman}%
  \BibitemOpen
  \bibfield  {author} {\bibinfo {author} {\bibfnamefont {M.}~\bibnamefont
  {Karpov}}, \bibinfo {author} {\bibfnamefont {H.}~\bibnamefont {Guo}},
  \bibinfo {author} {\bibfnamefont {A.}~\bibnamefont {Kordts}}, \bibinfo
  {author} {\bibfnamefont {V.}~\bibnamefont {Brasch}}, \bibinfo {author}
  {\bibfnamefont {M.~H.~P.}\ \bibnamefont {Pfeiffer}}, \bibinfo {author}
  {\bibfnamefont {M.}~\bibnamefont {Zervas}}, \bibinfo {author} {\bibfnamefont
  {M.}~\bibnamefont {Geiselmann}}, \ and\ \bibinfo {author} {\bibfnamefont
  {T.~J.}\ \bibnamefont {Kippenberg}},\ }\href {\doibase
  10.1103/PhysRevLett.116.103902} {\bibfield  {journal} {\bibinfo  {journal}
  {Phys. Rev. Lett.}\ }\textbf {\bibinfo {volume} {116}},\ \bibinfo {pages}
  {103902} (\bibinfo {year} {2016})}\BibitemShut {NoStop}%
\bibitem [{\citenamefont {Yi}\ \emph {et~al.}(2016{\natexlab{b}})\citenamefont
  {Yi}, \citenamefont {Yang}, \citenamefont {Yang},\ and\ \citenamefont
  {Vahala}}]{yi2016theory}%
  \BibitemOpen
  \bibfield  {author} {\bibinfo {author} {\bibfnamefont {X.}~\bibnamefont
  {Yi}}, \bibinfo {author} {\bibfnamefont {Q.-F.}\ \bibnamefont {Yang}},
  \bibinfo {author} {\bibfnamefont {K.-Y.}\ \bibnamefont {Yang}}, \ and\
  \bibinfo {author} {\bibfnamefont {K.~J.}\ \bibnamefont {Vahala}},\ }\href
  {\doibase 10.1364/OL.41.003419} {\bibfield  {journal} {\bibinfo  {journal}
  {Opt. Lett.}\ }\textbf {\bibinfo {volume} {41}},\ \bibinfo {pages} {3419}
  (\bibinfo {year} {2016}{\natexlab{b}})}\BibitemShut {NoStop}%
\bibitem [{\citenamefont {D'Aguanno}\ and\ \citenamefont
  {Menyuk}(2016)}]{daguanno2016nonlinear}%
  \BibitemOpen
  \bibfield  {author} {\bibinfo {author} {\bibfnamefont {G.}~\bibnamefont
  {D'Aguanno}}\ and\ \bibinfo {author} {\bibfnamefont {C.~R.}\ \bibnamefont
  {Menyuk}},\ }\href {\doibase 10.1103/PhysRevA.93.043820} {\bibfield
  {journal} {\bibinfo  {journal} {Phys. Rev. A}\ }\textbf {\bibinfo {volume}
  {93}},\ \bibinfo {pages} {043820} (\bibinfo {year} {2016})}\BibitemShut
  {NoStop}%
\bibitem [{\citenamefont {Haus}\ and\ \citenamefont
  {Huang}(1991)}]{haus1991coupled}%
  \BibitemOpen
  \bibfield  {author} {\bibinfo {author} {\bibfnamefont {H.~A.}\ \bibnamefont
  {Haus}}\ and\ \bibinfo {author} {\bibfnamefont {W.}~\bibnamefont {Huang}},\
  }\href {\doibase 10.1109/5.104225} {\bibfield  {journal} {\bibinfo  {journal}
  {Proceedings of the IEEE}\ }\textbf {\bibinfo {volume} {79}},\ \bibinfo
  {pages} {1505} (\bibinfo {year} {1991})}\BibitemShut {NoStop}%
\bibitem [{\citenamefont {Wiersig}(2006)}]{wiersig2006formation}%
  \BibitemOpen
  \bibfield  {author} {\bibinfo {author} {\bibfnamefont {J.}~\bibnamefont
  {Wiersig}},\ }\href {\doibase 10.1103/PhysRevLett.97.253901} {\bibfield
  {journal} {\bibinfo  {journal} {Phys. Rev. Lett.}\ }\textbf {\bibinfo
  {volume} {97}},\ \bibinfo {pages} {253901} (\bibinfo {year}
  {2006})}\BibitemShut {NoStop}%
\bibitem [{\citenamefont {Coen}\ and\ \citenamefont
  {Erkintalo}(2013)}]{Coen2013scaling}%
  \BibitemOpen
  \bibfield  {author} {\bibinfo {author} {\bibfnamefont {S.}~\bibnamefont
  {Coen}}\ and\ \bibinfo {author} {\bibfnamefont {M.}~\bibnamefont
  {Erkintalo}},\ }\href {\doibase 10.1364/OL.38.001790} {\bibfield  {journal}
  {\bibinfo  {journal} {Optics letters}\ }\textbf {\bibinfo {volume} {38}},\
  \bibinfo {pages} {1790} (\bibinfo {year} {2013})}\BibitemShut {NoStop}%
\end{thebibliography}

%

\end{document}